\documentstyle[11pt,aaspp4]{article}

\newcommand\gsim{{{}_>\atop{}^{{}^\sim}}}

\lefthead{Martini et al.}
\righthead{V4332 Sgr}

\begin{document}

\begin{center}
To appear in {\it The Astronomical Journal}, August 1999
\end{center}

\title{ Nova Sagittarii 1994 \#1 (V4332 Sagittarii): The Discovery and 
Evolution of an Unusual Luminous Red Variable Star}

\author{Paul Martini, R. Mark Wagner\altaffilmark{1} }
\affil{Department of Astronomy, Ohio State University, 
174 West 18th Ave., Columbus, OH 43210 \\ 
martini@astronomy.ohio-state.edu, rmw@as.arizona.edu}

\author{Austin Tomaney}
\affil{Department of Astronomy, PO Box 351580, University of Washington
Seattle, WA 98195}

\author{R. Michael Rich}
\affil{Department of Physics and Astronomy, UCLA, 405 Hilgard Ave., 
Los Angeles, CA 90095-1547}

\author{M. Della Valle}
\affil{Dipartimento di Astronomia, Vicolo dell'Osservatorio, 5, I-35122 
Padova, Italy}

\and

\author{Peter H. Hauschildt}
\affil{Department of Physics and Astronomy and 
Center for Simulational Physics, 
University of Georgia,
Athens, GA 30602-2451}

\altaffiltext{1}{Postal Address: Steward Observatory, University of Arizona, 
Tucson, AZ 85721}

\begin{abstract}

We report photometry and spectroscopy of the evolution of Nova Sagittarii 
1994~\#1 (V4332~Sagittarii) during outburst. We compare the photometric and 
spectral evolution of this outburst to known classes of outbursts --
including classical novae and outbursts occurring on symbiotic stars -- and 
find this object does {\it not} conform to any known class of outburst. 
The closest match to the behavior of this unusual object is M31~RV, an 
extremely luminous and red variable object discovered in the bulge of M31 in 
1988.  However, the temporal behavior and maximum luminosity of the two events 
differ by several orders of magnitude, requiring substantial intrinsic 
variation if these two events are members the same type of outburst. 

Our model of the spectroscopic evolution of this outburst shows that the 
effective temperature cooled from 4400 K to 2300 K over the three month span 
of our observations.  In combination with line diagnostics in our later 
spectra, including [OI] $\lambda 5577$ and the dramatic increase in the 
H$\alpha$ to H$\beta$ ratio, we infer the existence of a cool, dense 
($N_e \sim 10^{8-9}$ cm$^{-3}$) envelope that is optically thick in the 
Hydrogen Balmer recombination lines (case C).  We suggest that a nuclear event 
in a single star, in which a slow shock drove the photosphere outwards, can 
power the observed luminosity evolution and the emission spectrum. 

\end{abstract}


\section{Introduction}

On 1994 February 24, M. Yamamoto discovered an apparent nova in Sagittarius,  
Nova Sagittarii 1994 \#1 (hereafter V4332 Sgr; Hayashi \& Yamamoto 1994).  
A confirmation spectrum obtained by R. Bertram with the Perkins telescope on 
1994 March 4 indicated that this object was not a classical nova early in its 
outburst (Wagner 1994) since it lacked the spectral 
features characteristic of classical novae in outburst.
Subsequent spectra also showed this object to undergo approximately a one
magnitude decline in flux and a change in spectral type from approximately 
MO to M5 in only 5 days (Tomaney et al. 1994).  
Based upon its evolution towards later spectral type and the lack of any 
emission other than narrow Balmer lines, Tomaney et al. (1994) proposed that 
V4332 Sgr was of the same type as an extremely luminous red variable 
observed in the bulge of M31 (hereafter M31 RV) in 1988 (Rich et al. 1989; 
Mould et al. 1990; Sharov 1990; Tomaney \& Shafter 1992; Sharov 1993).  
M31 RV similarly exhibited only narrow Balmer emission, although this object 
evolved on a much longer timescale and attained a much brighter absolute 
magnitude than V4332 Sgr. 

In this paper, we compile all available observations of V4332 Sgr during 
outburst and show that its spectroscopic and photometric evolution are 
inconsistent with any known class of outburst.  In \S2, we describe the 
discovery and identification of V4332 Sgr. In \S\S 3 and 4 we trace its 
photometric and spectroscopic evolution, while in \S 5 we discuss our 
model fits to our spectra and derive effective temperatures for the expanding
shell as a function of time. In \S 6 we describe high-resolution spectroscopy 
of H$\alpha$ at two epochs, while in \S\S 7 and 8 we compare V4332 Sgr with 
other known classes of outbursts and M31 RV. We briefly discuss the possible 
nature of the object and this outburst in \S 9 and we summarize our results 
in \S 10. 

\section{Discovery, Identification \& Astrometry}

V4332 Sgr was discovered by Minoru Yamamoto on an exposure with P-MAX 400 
film on 1994 February 24.85 UT at a magnitude of 8.9 (Hayashi 
1994).  The presence of the new object was confirmed visually by Hirosawa 
(1994) on February 25.85 at a magnitude of about 8.4.  Further observations by 
Yamamoto (1994) on February 25.84 gave a magnitude estimate of 8.5, consistent 
with the results of Hirosawa.  V4332 Sgr should not be confused with a second 
nova in Sagittarius (Nova Sagittarii 1994 \#2) discovered by Sakurai (Sato 
1994) on 1994 May 20.7. 

In Figure~1, we show the location of V4332 Sgr based on a 10~s R-band CCD 
image obtained on 1994 March 22 using the Imaging Fabry-Perot 
Spectrometer (Pogge et al. 1995) in direct imaging mode and a 
TI $800 \times 800$ pixel CCD mounted on the 
Perkins telescope of the Ohio Wesleyan and Ohio State Universities at the 
Lowell Observatory.  Figure~1 covers a region $6\farcm0 \times 6\farcm3$ at a 
scale of 0\farcs5 per pixel. The seeing was approximately $2\arcsec$ FWHM.  

Kilmartin (1994) obtained astrometry of V4332 Sgr by utilizing the Parallax 
and Proper Motion Catalog (PPM) based on two exposures obtained by 
A. C. Gilmore with the 0.15 m f/15 astrograph at Mt. John Observatory on 
March 2.62 UT.  Kilmartin determined a mean FK5/J2000 position of 
$\alpha = 18^h 50^m 36\fs74, \delta = -21\arcdeg23\arcmin28\farcs8$.
Skiff (1994) examined the Palomar Observatory Sky Survey (POSS) O and E prints
and found a possible progenitor of the nova at this position with R = 16 mag 
and B = 18 mag at a mean position of $\alpha = 18^h 50^m 36\fs80, \delta = 
-21\arcdeg23\arcmin29\farcs3$ (J2000). With our estimated extinction of 
$A_V = 1$ (see below), this is roughly consistent with a K star. This is also
consistent with the initial spectral type of the outburst. 

From the R-band CCD images we obtained on 1994 March 22, we derived 
an astrometric solution for 27 stars which were in common with our CCD
image and the Digitized Sky Survey based upon a grid of 18 PPM stars
in a $100\arcmin \times 100\arcmin$ field centered on V4332 Sgr. Using
our plate solution, we obtained a position of $\alpha = 18^h 50^m 36\fs73,
\delta = -21\arcdeg23\arcmin28\farcs98$ (J2000) with a formal uncertainty
of 0\farcs26 in both coordinates. Our position is consistent at the
1$\sigma$ level with the position derived by Kilmartin (1994) and with the
position of the possible progenitor on the POSS print discovered by Skiff 
(1994).

\section{Light Curve}

In Figure~2a, we plot the light curve of V4332 Sgr as compiled by VSNET, the 
AAVSO, and photometry reported in the IAU Circulars (Nos. 5943, 5944, and 
5949). We show digital measurements obtained by photoelectric photometry 
or with CCDs as filled circles, while we denote visual magnitude estimates by 
open circles. Visual estimates were obtained and kindly communicated to us by 
Mr. Albert Jones of the Variable Star Section of the Royal Astronomical 
Society of New Zealand using a 0.32 m telescope. These points are shown as 
open triangles. The photographic observations, primarily by Sakurai (1994), 
largely precede the discovery observation and are shown as filled 
triangles.  For convenience, we have included the results of published 
photoelectric photometry in Table~1.

The light curve is characterized by a slow rise in brightness to a relatively 
flat maximum. Maximum brightness appears to have been reached on 
JD 2,449,412 $\pm 1$ (1994 February 28) at a visual magnitude of $\sim 8.5$.
This broad maximum was then followed by a relatively fast decay in which the 
time to decline by 2 and 3 magnitudes in $V$ was $t_2 = 8 \pm 1$ and $t_3 = 11 
\pm 1$ days, respectively.  In addition, the visual magnitude
measurements suggest that the rate of decline slowed after JD 2,449,424 
(1994 March 12). We performed a crude linear back-extrapolation of the
pre-maximum light curve data (filled triangles in Figure~2a) and 
estimate that the object took $\sim$200 days to reach maximum brightness, 
assuming a quiescent magnitude of $B \simeq 18$ mag based on Skiff's (1994)
proposed progenitor. Archival plates examined by Wenzel (1994) indicate 
that V4332 Sgr was not visible on 561 Sonneberg sky patrol plates taken 
between 1926 and 1983 to a limiting photographic magnitude of 11-13.5.  
We note that this object was just becoming visible in the morning sky when 
it was discovered and therefore more complex photometric behavior prior 
to its discovery can not be excluded. 

In Figure~2b, we show the $V-I$ color evolution taken from the photometry 
reported by Gilmore (IAUC Nos. 5943, 5944, and 5949) starting 
from maximum light and extending for 11 additional days (see also Table~1).  
During this time, the slope of the continuum of V4332 Sgr as measured by the 
$V-I$ color became redder by $\sim$ 1.5 mag while the object faded by 
2.5 mag in $V$. This suggests that much of the steep decline in the visual 
magnitude was the result of a rapidly reddening continuum. This behavior is 
confirmed by the spectral evolution of V4332 Sgr discussed below.

\section{Spectral Evolution}

We followed the early spectral evolution of V4332 Sgr from 1994 March 4 to 
June 6 at a variety of telescopes (Table~2). In Figure~3, we present spectra 
of V4332 Sgr obtained at the Perkins telescope on 1994 March 4.5 (Figure~3a) 
and 9.5 UT (Figure~3b) with the Boller and Chivens CCD spectrograph.  We 
employed a 350 grooves mm$^{-1}$ grating centered at 5100 \AA\ and a 2\arcsec\ 
wide slit to cover the spectral region from $4500-5700$ \AA\ at a resolution 
of 5 \AA.  Upon comparing this spectrum with the Turnshek et al.\ (1985) atlas 
of cool stars and the spectra presented by Jacoby, Hunter, \& Christian (1984) 
we concluded that the the March~4 spectrum (Figure~3a) most closely resembles 
a K3-4 III-I star.  In addition to the stellar absorption features, a narrow 
and unresolved H$\beta$ emission line with an equivalent width of 3 \AA\ is 
present. The March~9 spectrum (Figure~3b), only 5 days later, shows the rapid 
spectral evolution of V4332 Sgr.  During this interval, the spectral type 
changed from that of a K3-4 III-I star to that of a M3 III-I star, although 
atmospheric model fitting (see \S 5 below) yields an effective temperature 
consistent with a K5 III.  This discrepancy is probably due to the limited 
wavelength range of this spectrum (see Table~2).  Over this same 5-day period 
the $V-$band brightness decreased by approximately one magnitude.  Unresolved 
H$\beta$ emission is present in this spectrum with an equivalent width of 
11.4 \AA.  The TiO absorption features which were weak or absent on March 4 
are now conspicuous in the spectrum.  Without more detailed spectra at higher 
spectral resolution in 
the blue or the near infrared it is difficult to distinguish between 
luminosity classes of III and I at this spectral type.  In any event, using 
luminosity classifications meant for static atmospheres on an outburst are 
likely to be very uncertain. 

Beginning on March 11, we were able to secure spectra which covered a much 
larger portion of the visible-wavelength region than our previous observations.  
In Figure~4a, we show the spectrum of V4332 Sgr obtained on 1994 March 11.4 
with the 3.6 m telescope at La Silla using EFOSC.  The spectrum covers the 
region $3800-8600$ \AA\ at a resolution of 7 \AA\ and was obtained under 
non--photometric conditions. Comparison with M--star templates from the sources 
cited above suggests that the spectrum is that of a M5-6 III or M5-6 Ib-II 
star. At our resolution, it is difficult to distinguish between these 
luminosity classes.  Emission lines arising from the Balmer series are now 
striking over our extended spectral range.  The observed ratios of the 
H$\alpha$, H$\gamma$, and H$\delta$ emission lines to H$\beta$ are 2.6, 0.8, 
and 0.8, respectively, with a formal measurement uncertainty of 10\%, though 
we note that there is significant additional uncertainty in estimating the 
true level of the underlying stellar continuum as well as absorption from 
the photosphere.  In particular, our high-resolution spectra (see below) 
show absorption in the H$\alpha$ profile, so clearly these line ratios are 
very approximate.  These caveats apply to all of our reported line ratios. 

On March 20.4, we obtained another spectrum with the 3.6 m telescope and EFOSC 
under photometric conditions at a resolution of 13 \AA\ and covering the range 
$3750 - 9800$ \AA\ (Figure~4b).  This spectrum is more typical of a M6.5-7 III 
or II star since the VO Bands at $\lambda\lambda$ 7400 \AA\ and 7900 \AA\ are 
now discernible.  The observed ratios of the H$\alpha$, H$\gamma$, and 
H$\delta$ emission lines to H$\beta$ are 7.5, 0.6, and 0.7, respectively, 
with a formal measurement uncertainty of 10\%.  Weak emission due to 
Mg~I at 4571 \AA\ is also visible in both spectra, although no forbidden lines
are apparent.

In Figure~5 we present a composite spectrum of V4332 Sgr obtained on 1994 
June 5 \& 6 using the 4.5-m Multiple Mirror Telescope (MMT) and the Blue 
Channel CCD spectrograph.  The June 5 spectrum covers the region $3650 - 7200$ 
\AA\ and the June 6 spectrum covers the region $5600 - 9100$ \AA, both at a 
resolution of 3.5 \AA.  The spectrum continued to evolve strongly towards 
later spectral types and decrease overall in absolute flux from early March, 
although we cannot infer a $V$ magnitude from the spectrum due to slit 
losses.  From the strength of the VO bands, we classify this spectrum as an 
M8-9 III.  In 
addition to the narrow Balmer emission lines present in earlier spectra, 
emission lines arising from Mg~I, Na I, [O I], Fe I, and Fe II are also 
present.  The observed ratios of the H$\alpha$, H$\gamma$, and H$\delta$ 
emission lines to H$\beta$ are 18.6, 0.26, and 0.12 respectively, again with 
a formal measurement uncertainty of 10\%.  The equivalent widths of the Balmer 
series and other prominent lines are listed in Table~3.  

Figure~6 shows the color excess evolution of V4332 Sgr for 3 dates on which 
we obtained spectra and also have relatively coincident color information. The 
points plotted in this figure represent the observed color excess of V4332 Sgr 
which we derived from the difference between the published photoelectric
photometry in Table~1 and the nominal color of giant stars (Johnson 1966) of 
our observed spectral type. The errorbars represent an assumed uncertainty of
1 spectral type. This figure shows that the average $E(B-V)$ is relatively 
constant and consistent with the value we infer for March 4, $E(B-V) 
\sim 0.32 \pm 0.02$. If this component is due to interstellar reddening, 
then $A_V \simeq 1$ mag (assuming $R = 3.1$).  Both $E(V-R)$ and $E(V-I)$ 
appear to be somewhat bluer than the assumed colors for static atmospheres; 
this may be due to a poor match between the inferred temperatures and 
static atmosphere colors or scattering in the expanding envelope. The effect 
seen in Figure~6 is in the opposite sense of what one would expect if dust 
were forming in the envelope.

\section{Modeling of the Spectra}

We have modeled the spectra using the {\tt PHOENIX} model atmosphere code
(Hauschildt \& Baron 1998; Hauschildt, Allard, \& Baron 1998).  The presence
of molecular bands in the observed spectra indicates low effective 
temperatures.  Therefore, we have calculated models using essentially the same 
input physics as described in Hauschildt, Allard, \& Baron (1998).  However, we 
have used spherical (rather than plane parallel) geometry for the model 
construction due to the geometrical extension of the changing atmosphere.  In 
addition, the models presented here were calculated using the AMES water line 
list (300 million lines: Partridge \& Schwenke 1997) which replace the older 
water data used in Hauschildt, Allard, \& Baron (1998).  For this preliminary 
modeling we have neglected the small expansion velocity (see below) of the 
shell and used static atmosphere models in order to save computational time.  
The formation of the molecules and their lines were treated in LTE (molecular 
NLTE models are in preparation but have not yet been finalized) and all models 
were calculated assuming solar abundances.  Before the modeling, the observed 
spectra were corrected for interstellar reddening assuming $E(B-V) = 0.32$ as 
derived above.

The largest difference between the spectra of V4332 Sgr and that of a 
``normal'' classical nova atmosphere early in its outburst (the ``iron curtain 
phase'') is the significantly lower effective temperature.  The emission 
lines that are apparent in the observed spectra are true emission features, 
evident by a comparison of the model spectra with the observed data.  The 
models do not include a circumstellar shell where these lines probably form 
and thus cannot be expected to reproduce them.  Our spectra do however 
resemble active M-type spectra (the gravities in the dMe's are larger 
however) more than classical nova spectra.  In a statistical sense, the model 
spectra do not 
reproduce the observed spectra very well.  In part this is due to the model 
simplifications (abundances, LTE). Another major problem is the incomplete
line list data for TiO we had to use in these models (Hauschildt et al, in
preparation), as well as unknown line data for molecules like VO and FeH.
However, we can say that the effective temperature of the spectra dropped
throughout the observed time period by nearly a factor of 2 while the spectra
continued to stay optically {\em thick}.  This is in sharp contrast to normal
novae, where the effective temperature of the shell increases and it becomes
optically thin. 

Our models fits to the March~4 and March~9 spectra show a decline in 
temperature from $\sim 4400$~K to 3800~K.  While the March~4 fit is in good 
agreement with our spectral classification of K3 (see Figure 7), our March~9 
fit yields a hotter temperature than our spectral classification.  This 
inconsistency is probably due to the narrow observed spectra range on 
this data.  The model fits to March~11 ($\sim 3100$~K), March~20 
($\sim 2600$~K), and June~5/6 ($\sim 2300$~K) are all consistent with the 
spectral classification discussed above.  Our derived effective temperatures 
for each of the spectra are listed in Table~2.  In Figure~7 we show the 
observed spectra ({\it bold tracing}) along with the model fits ({\it thin 
tracing}).  

The static atmosphere models we have used to obtain temperature estimates 
also provide us with a way to compare the change in apparent $V$ magnitude 
to the change in the bolometric luminosity.  As shown on Table~1, V4332~Sgr 
decreases by approximately 10 magnitudes in $V$ over the 3 month period from 
the beginning of March to the beginning of June, over the same period when 
our model fits show a decrease in effective temperature from 4400~K to 
2300~K.  When we use the bolometric corrections of these models to calculate 
the change in the bolometric luminosity over this time, we find that the 
bolometric luminosity decreases by a factor of $\sim 100$ over this 3 month 
period. 

\section{Line Profiles}

In order to learn more about the dynamics of the cool, evolving envelope 
indicated by our low dispersion spectroscopy, we obtained a higher dispersion 
spectrum on 1994 March 7 in the H$\alpha$ region using the 3.5 m New 
Technology Telescope (NTT) at La Silla with EMMI.  Using a 1\farcs1 wide 
entrance slit and a 1200 grooves mm$^{-1}$ grism, we obtained 
a spectrum covering the range $6200 - 6800$ \AA\ at a resolution of 1.4 \AA. 
The region from $6545 - 6575$ \AA\ is shown in Figure~7a in which the 
continuum normalized H$\alpha$ profile is represented by crosses.  
The line profile consists of a narrow emission line blended with a redshifted 
but narrower absorption line which together form an inverse P Cygni profile.

To extract quantitative information from this profile we deconvolved the two 
components by Gaussian fitting with a non--linear least squares technique.  
Our best--fit is shown as a solid line superposed on the observations in 
Figure~7a.  We find that the absorption component is redshifted with respect 
to the emission component by 17 km s$^{-1}$.  The FWHM of H$\alpha$ emission is 
200 km s$^{-1}$ while that of H$\alpha$ absorption is 70 km s$^{-1}$, both 
corrected for instrumental resolution.  These results suggest that the 
H$\alpha$ line emitting region is expanding with a velocity of about 100 km 
s$^{-1}$.  If the emission 
and absorption components are moving with respect to the same center of mass, 
an infall velocity of about 20 km s$^{-1}$ is required of a component of the 
envelope projected against the continuum source.  Another striking aspect of 
Figure~7a is that the systemic velocity of the mean H$\alpha$ profile is 
blueshifted with respect to line center by 180 km s$^{-1}$.  This blueshift is 
also evident in our other spectra (see, e.g. Table 3) and may be due to an 
occulting source in or near to the emission line region, the effects of an 
opacity source in the envelope such as dust obscuring the far side of the 
emitting envelope, the actual heliocentric radial velocity of V4332 Sgr, or
a combination of all of these effects.  

On 1994 April 26 we obtained a second high resolution spectrum of V4332 Sgr 
using the Palomar Observatory 1.5-m telescope and echelle spectrograph at a 
resolution of 1 \AA.  The spectrum covers the region from $3670-9960$ \AA\ in 
58 orders. We were unable to perform a sky subtraction or flux calibration for 
this spectrum as we lacked the appropriate calibration data.  The emission 
lines we identified in the spectrum include the Balmer series of hydrogen 
and emission lines arising from Ca I 4226 \AA, Ca II H \& K lines, the 
Ca II infrared triplet, O I 8446 \AA, weak [O I] 6300 \AA\ and 6363 \AA, 
Na I D, and Mn I multiplets 1 and 2 at 5394 \AA, 5432 \AA, and 4030 \AA\ 
and numerous emission lines from Fe I and Fe II.  The lack of sky subtraction 
makes positive identifications of other emission lines ambiguous.  All of 
the identified emission lines exhibit the blueward shift described above. 
The [O I] detections are clearly not the telluric features as they are 
shifted blueward of the telluric lines.  
In Figure~7b, we show the H$\alpha$ line profile obtained on April 26.  
The same basic structure is present in the line profile at this epoch as 
observed on March 7 except that the relative emission and absorption line 
strengths have changed and the line widths are narrower. 

\section{Comparison with Known Outbursts and Variable Stars}

This object was referred to as a nova before the identification spectrum 
obtained by Wagner (1994) showed that its spectrum was completely inconsistent
with this classification. The collection of observations we present here 
show that this object also had a drastically different ejection velocity and 
temporal evolution from classical novae (Gallagher \& Starrfield 1978).  
Since novae do not fit the characteristics of this event, we investigated 
the possiblility that this object could be another type of previously cataloged 
outburst.  Symbiotic stars, for example, constitute a fairly diverse class of 
objects which exhibit a range of outburst phenomena (Kenyon 1986).  The 
different classes of outburst exhibited by symbiotic stars, however, all 
include high ionization emission lines and have much lower amplitudes than the 
observed $\gsim 10$ magnitude variation in V4332 Sgr.  Finally, the relatively 
rapid luminosity evolution of this object, lack of evidence for periodicity, 
and the different spectral features make it unlikely that this object is an 
asymptotic giant branch star undergoing a final helium shell flash, such as 
FK~Sgr. 

\section{Comparison with M31 RV}

From the above comparison with known types of outbursts and variable stars, 
we find that V4332 Sgr does not conform to the criteria for membership in any 
of these classes.  What is particularly striking is its unique spectral 
development from K3 III-I to M6 III-I in just one week and subsequently to 
M8-9 III in less than 3 months.  Only the extremely luminous outburst in M31 
in 1988, M31 RV (Rich et al. 1989; Mould et al. 1990), exhibited similar 
spectral evolution.  Both M31 RV and V4332 Sgr were characterized by a rapidly 
evolving late-type stellar absorption line spectrum and narrow Balmer emission 
lines with a rapidly increasing H$\alpha$ to H$\beta$ ratio with time.  Both 
outbursts were not characteristic of classical novae, symbiotic stars, other 
known types of outbursts, or known variable stars and neither have exhibited 
evidence of periodicity.  In addition, the spectra of both objects exhibited 
no other emission lines than narrow Balmer emission lines until at 
least $t_2$ days after maximum.

M31 RV was discovered as a 15th mag red variable star 3\arcmin\ from the 
center of M31 on 1988 September 3 by Rich et al. (1989) during the course of 
a survey of luminous M giants in the bulge of M31.  It was independently 
discovered by Tomaney \& Shafter (1992) on 1988 August 18 during their study 
of novae in the bulge of M31 and by Bryan \& Royer (1992) on 1988 July 13 
during their photographic survey for M31 novae at the Ford Observatory.  The 
time of the outburst was constrained by Bryan \& Royer (1992) to have taken 
place sometime between 1988 June 14 and July 13 based on independent 
photographs of M31 obtained by A. N. Sollee (1990).  

Spectroscopy of M31 RV obtained by Rich et al. (1989) on 1988 September 5 
exhibited unresolved H$\alpha$ and H$\beta$ emission, Na absorption, weak 
TiO bands, and a strong Ca II IR triplet.  Based on the strength of the Ca II
triplet, the lack of CaH absorption, and the strength of the $\lambda 6500$ 
blend (Ba II, Ca I, and Fe I absorption blend), Rich et al. classified it as 
M0 Ie.  Additional spectra (Mould et al. 1990) were obtained in early October 
and November 1988 and indicated that the spectrum of M31 RV had changed to 
later M types.  By 1988 Nov 1, the spectrum was classified as M6 Ie.  No 
emission lines other than the Balmer series were reported.  Rich et al. 
reported dramatic variations in its radial velocity in their September 5 
spectra in both the emission and absorption lines.  Subsequent spectra 
obtained by Mould et al. (1990) gave a blueshift relative to the systemic 
velocity of M31 of $\simeq150$ km s$^{-1}$.

The light curves of M31 RV and V4332 Sgr are very different in 
their relative rates of decline after maximum brightness.  As measured by the 
time to decline 2 mag in brightness from maximum, the rate of decline for 
V4332 Sgr was $t_2 = 8$ days whereas for M31 RV $t_2 =$ 78 days.  We note, 
however, that the time of maximum of both of these objects is somewhat 
uncertain. The maximum of M31 RV has been constrained by observation to only 
within a few weeks (Bryan \& Royer 1991; Sharov 1993). The time of maximum 
brightness of V4332 Sgr could be even more uncertain as it could have reached 
maximum prior to its discovery.  In either scenario, this would impact upon 
the value of $t_2$ used to characterize the temporal evolution of these 
objects. 

Another difference between M31 RV and V4332 Sgr is the maximum absolute 
magnitude achieved during their outbursts.  M31 RV reached an apparent 
magnitude of $\simeq15$ in 1988 July.  Assuming $(m-M)=24.2$ mag for M31, the 
bolometric absolute magnitude was $M_{bol} = -10$ (Rich et al. 1989).  While 
we do not know the distance to V4332 Sgr, there are many lines of evidence, 
such as the relatively weak interstellar absorption lines and low extinction 
estimate, which suggest that the progenitor was relatively nearby and similar 
to a late-type dwarf or giant star, rather than a late-type supergiant.  
If we assume V4332 Sgr was characteristic of a K giant at maximum, as 
suggested by Figure~3, the implied distance is $\sim 300$ pc, before including 
the effects of extinction.  Performing the same exercise for subsequent points 
in the evolution leads to increasingly larger distance estimates as the object 
evolves towards later spectral types.  The luminosity of this object is 
therefore not consistent with the observed spectral type and luminosity class, 
though as the luminosity increases with time over that expected for a star of 
the same spectral type, the distance estimate of $\sim 300$ pc derived from 
the first spectrum may be an upper limit.  We note as above that this line of 
analysis is very uncertain as we do not expect the luminosity class of the 
object during outburst to provide a reasonable luminosity estimate in any 
event as these luminosity classifications are based on static stellar 
atmosphere. 

There is a spectroscopic similarity between V4332 Sgr and M31 RV in the strong 
increase in 
the H$\alpha/H\beta$ intensity ratio, possibly indicative of greatly 
increasing reddening at the source such as that produced by dust formation.
V4332 Sgr had a H$\alpha/H\beta$ intensity ratio of 2.5 on March 11
($t_{dis} +$ 15 days, where $t_{dis}$ is the date of discovery: 1994 Feb
24). By June 5 ($t_{dis} +$ 40 days) this ratio had increased to $\sim 17$.
Mould et al. (1990) reported M31 RV to have an H$\alpha/H\beta$ emission
line ratio of 1.8 at $t_{dis} + 82$ days. While by $t_{dis} +$ 111 to 119 
days this ratio had increased to 9.8 (Tomaney \& Shafter 1992). Both Mould
et al. (1990) and Tomaney \& Shafter (1992) interpret the continuum
evolution and increase in the H$\alpha/H\beta$ emission line 
ratio as due to dust formation. However, as we discuss below, we attribute 
this change to increasing optical depth in the Hydrogen Balmer recombination 
lines. 

\section{Discussion}

Both M31 RV and V4332 Sgr do not conform to any known class of outburst. 
While M31 RV exhibited temporal and luminosity evolution orders of magnitude
greater than V4332 Sgr, they may be related based upon their similar spectral 
evolution.  M31 RV and V4332 Sgr evolved rapidly towards the red and exhibited 
an increasing H$\alpha$ to H$\beta$ line emission ratio with time and thus it 
is possible that both outbursts produced a great deal of dust.  However, the 
behavior of Figure~6 is inconsistent with this interpretation for V4332 Sgr.
Except for a minor trend in the color excesses of $V-R$ and $V-I$ with time
which might be due to a systematic error with increasing spectral type in the
intrinsic colors of M giant stars, there is no indication that dust formation
is taking place and resulting in large color excesses that would lead to
large $H\alpha/H\beta$ intensity ratios. The observed colors are in good 
agreement with the temperatures derived from our atmosphere model without the 
addition of dust beyond our estimate of an interstellar extinction of 
$A_V = 1$. Instead, large H$\alpha/$H$\beta$ intensity ratios can be produced 
if the Balmer lines become optically thick (case C recombination).  This 
situation would then be similar 
to that observed in the evolution of the visible-wavelength spectra of type II 
supernovae.  In their model for hydrogen recombination in SN 1987A, Xu et al. 
(1992) found large values for the H$\alpha$ to H$\beta$ ratio in their 
$N_e = 10^8$ cm$^{-3}$, $T = 3000$ K model for the late-time spectra. 

We have independent evidence for such high densities and low temperatures in 
our spectra which further support these values for the physical parameters.  
As discussed in section 5, our models show that the temperature changes 
from $T \sim 4400$ K to $T \sim 2300$ K 
over the three month period we have spectra.  In addition, we detect [O I] 
$\lambda 5577$ \AA\ in our June 1994 spectrum.  The ratio of this line to the 
$\lambda 6300$ \AA\ doublet is approximately 0.08, implying $N_e \sim 10^{8-9}$ 
cm$^{-3}$ (Begelman \& Sarazin 1986; Keenan et al. 1995).  The 
ratio of the [O I] $\lambda 6300$ \AA\ doublet is $\sim 3$, however, indicating 
the [O I] emission is optically thin.  In addition, the absence of forbidden 
line emission from either outburst within $t_2$ of maximum suggests a high 
density environment.  The critical densities of bright, forbidden emission 
lines typically seen in outbursts are generally $10^{4-7}$ cm$^{-3}$ and the 
absence of any of these lines provides a lower limit to the density of the 
line-emitting region. 

Iben \& Tutkov (1992) suggested that M31 RV could be a nova-like outburst 
involving an uncharacteristically large amount of mass loss ($\sim 10^{-2} 
M_{\sun}$).  In their scenario, M31 RV could be the first outburst to occur 
in a short period (80 minutes to 2 hours) cataclysmic variable system, where
the high mass loss and large luminosity would then be due to an exceptionally
large mass build-up on a cold white dwarf. This large amount of mass loss 
could also explain the lack of a nebular phase as the large amount of 
ejected mass may be absorbing all of the ionizing photons from the central 
source.  Our spectral and dynamical analysis of V4332 Sgr is not inconsistent 
with the Iben \& Tutkov (1992) model for M31 RV except for the lack of 
evidence for dust formation. However, further theoretical modeling is needed 
to see if such an outburst on a cold white dwarf could also produce the much 
lower maximum luminosity that V4332 Sgr exhibits.  Such a great intrinsic 
variation in `first outburst' events could be explained, for instance, by 
highly variable mass-loss and the proportion of energy that is emitted in the 
form of kinetic energy vs. radiation. 

An alternative explanation for this event and M31 RV could be a nuclear event 
in a single, evolved star which caused a slow shock to propagate through 
to the photosphere.  This would push out the photosphere of the star, 
increasing the luminosity and correspondingly decreasing the effective 
temperature.  This scenario explains the high densities, low velocities, 
and timescale we observe.  This is also provides a means of powering the 
event in the absence of any evidence for a $UV$ source or that this object 
is in a binary system.  Such evidence would consist of radial velocity 
variations, composite spectra, or evidence for a companion as the object 
faded, none of which have been observed.  Our observations are also consistent 
with the ejection of a massive and opaque shell of material. 

\section{Summary} 

We have presented photometry and spectroscopy that trace the evolution
of V4332 Sgr, an unusual luminous red variable star in Sagittarius.  Our 
data clearly rule out the possibility that this object was a nova or any 
other known type of variable object.  In fact the closest match to the 
spectral evolution of this object is M31 RV, an extremely luminous and red 
variable that was observed in the bulge of M31 in 1988. However, significant 
differences exist between these two events.  Particularly, the temporal 
evolution was much longer and the maximum luminosity much brighter for 
M31 RV.  Therefore if these two events are similar, there must be substantial 
intrinsic variation in the outbursts. 

Our model of the spectra observed over a 3 month baseline show that the 
effective temperature fell from $\sim 4400$~K to 2300 K.  Our line diagnostic 
measurements in the later spectra, particularly the evolution of the 
H$\alpha$ to H$\beta$ ratio and the absence of strong forbidden lines, imply 
high ($N_e = 10^{8-9}$ cm$^{-3}$) densities in addition to the low 
temperatures.  Our observations of this object suggest a simple model in 
which a nuclear event in a single, evolved star caused a slow shock to drive 
the photosphere outwards and resulted in the evolution to lower temperatures 
and the emission spectrum we see after three months.  Further study of this 
object in quiescence as well as theoretical modeling based upon the 
observational data we present here will hopefully lead to a clearer 
understanding of the nature of this unusual outburst. 

\acknowledgments

PM would like to thank Lowell Observatory for their hospitality while much of 
this work was completed. We would also like to acknowledge the support staff 
of Lowell Observatory, the NTT, La Silla, Palomar, and the MMT, where these 
observations were obtained. PHH acknowledges partial support by NASA ATP grant 
NAG 5-3018 and LTSA grant NAG 5-3619 and NSF grant AST-9720804 to the 
University of Georgia.  Some of the calculations presented in this paper were 
performed on the IBM SP2 and SGI Origin 2000 of the UGA UCNS, at the San Diego 
Supercomputer Center (SDSC), and at the National Center for
Supercomputing Applications (NCSA), with support from the National Science
Foundation, and at the NERSC with support from the DoE. We thank all these
institutions for a generous allocation of computer time.  Thanks go to Phil 
Pinto, Anil Pradham, and Alison Sills for several helpful suggestions and 
Darren DePoy and the referee, Steve Shore, for many useful comments on the 
manuscript.

\begin{deluxetable}{lccccc}
\tablenum{1}
\tablewidth{500pt}
\tablecaption{Photometric Evolution}
\tablehead {
\colhead{Date of Observation (UT)} &
\colhead{$V$} &
\colhead{$U-B$} &
\colhead{$B-V$} &
\colhead{$V-R$} &
\colhead{$V-I$} \\
}
\startdata
1994 March  3.67 &   8.67 &  +1.22 & +1.65 & +0.93 & +1.82  \nl
1994 March  4.67 &   8.76 &  +1.4  & +1.73 & +0.96 & +1.91  \nl
1994 March  7.71 &   9.82 &        & +1.97 & +1.07 & +2.27  \nl
1994 March  8.70 &  10.29 &        & +1.94 & +1.20 & +2.59  \nl
1994 March 10.71 &  11.18 &        & +1.85 & +1.32 & +3.09  \nl
1994 March 11.4  &  12.56 &        & +1.63 & +2.01 &        \nl
1994 June  5/6   &  18.78 &        & +0.90 & +4.12 &        \nl
\enddata
\tablecomments{
Photometry of Nova Sagittarii 94 by A.C. Gilmore with
the 0.6-m f/16 Cassegrain at Mt John Observatory. Typical uncertainties
are 0.05 magnitude, and 0.1 in U-B (1$\sigma$). The photometry listed for
March 11 and June 5 was obtained by convolving the March 11 (Figure~4) and
June 5 (Figure~5) spectra with standard filters.  These measurements were
then placed on an absolute scale by interpolating the
photoelectric V values shown in Figure~2. We estimate the uncertainties
to be approximately 10\%.
}
\end{deluxetable}

\begin{deluxetable}{lcclc}
\tablenum{2}
\tablewidth{500pt}
\tablecaption{Spectroscopic Observations and Model Results}
\tablehead {
\colhead{Date (UT)} &
\colhead{Spectral Coverage (\AA)} &
\colhead{Resolution (\AA)} &
\colhead{Telescope} &
\colhead{Model Temperature (Kelvin)} \\
}
\startdata
1994 March 4.5  & $4500 - 5700$ & 10    & 1.8-m Perkins & 4400 \nl
1994 March 7.   & $6200 - 6800$ & 1.4   & 3.5-m NTT     & \nl
1994 March 9.5  & $4500 - 5700$ & 10    & 1.8-m Perkins & 3800 \nl
1994 March 11   & $3800 - 8600$ & 7     & 3.6-m La Silla& 3100 \nl
1994 March 20.4 & $3750 - 9800$ & 13    & 3.6-m La Silla& 2600 \nl
1994 April 26   & $3670 - 9960$ & 0.5   & 1.5-m Palomar &  \nl
1994 June 5     & $3650 - 9100$ & 3.5   & 4.5-m MMT     & 2300 \nl
\enddata
\tablecomments{
The log of our spectroscopic observations of V4332 Sgr.
Column 1 shows the date of observation, columns 2 \& 3 the spectral
range and resolution of the data, respectively, and column 4 the telescope
at which we obtained the data (see \S 4). Column 5 shows the effective
temperature we have derived for each spectrum using the {\tt PHOENIX} model
atmosphere code (see \S 5).
}
\end{deluxetable}

\begin{deluxetable}{ccc}
\tablenum{3}
\tablewidth{500pt}
\tablecaption{Line Identifications from 1994 June 5/6}
\tablehead {
\colhead{Observed Wavelength (\AA)} &
\colhead{Line Identification(s)} &
\colhead{Equivalent Width (\AA)} \\
}
\startdata
3889.3  & Fe I $\lambda 3887.1$ + H$_8$ & 62    \nl
3968.1  & H$\epsilon$                   & 32    \nl
4099.8  & H$\delta$                     & 115   \nl
4200.3  & Fe I $\lambda 4202.0 $        & 65    \nl
4305.8  & Fe I $\lambda 4307.9$         & 30    \nl
4338.3  & H$\gamma$                     & 102   \nl
4413.5  & [Fe II] $\lambda 4416.3 $     & 36    \nl
4569.0  & Mg I $\lambda 4571.1$         & 172   \nl
4858.3  & H$\beta$                      & 305   \nl
5337.9  & Fe I $\lambda 5341.0  $       & 90    \nl
5889.7B & Na I $\lambda\lambda 5890.0 $ &       \nl
        & $+$ Na I $\lambda 5895.9 $    & 40    \nl
6297.2  & [O I] $\lambda 6300.2$        & 70    \nl
6360.6  & [O I] $\lambda 6363.9$        & 41    \nl
6558.7  & H$\alpha$                     & 450   \nl
\enddata
\tablecomments{
The identifications of the brightest lines in the
1994 June 5/6 spectrum shown as Figure 5. Column 1 lists the observed
wavelength of the emission feature, while column 2 lists the identification
of the line(s).  The equivalent widths of the features are given in column 3.
}
\end{deluxetable}

\begin{figure}
\plotfiddle{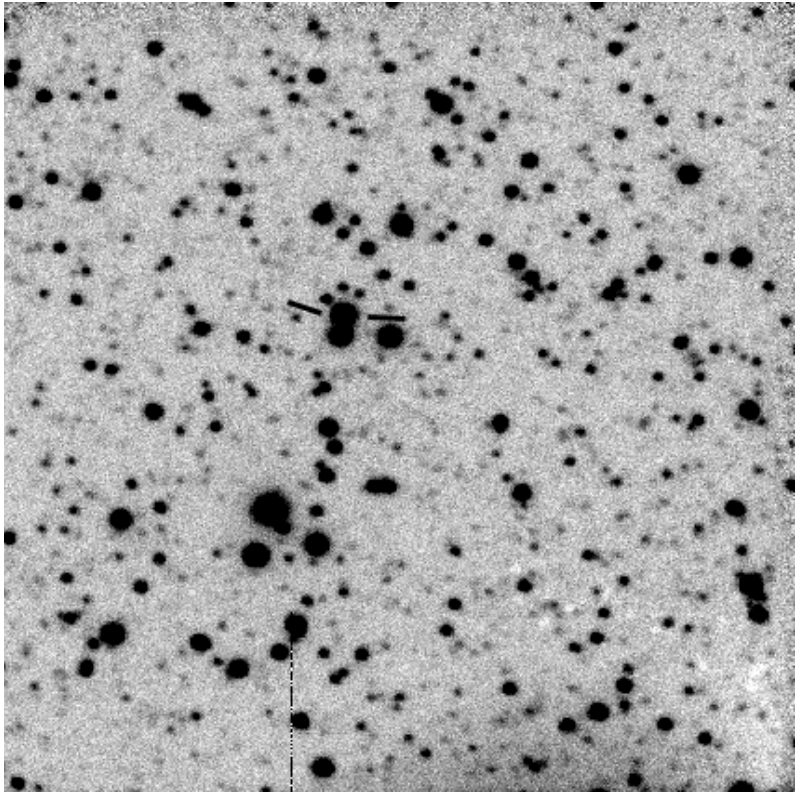}{5.0truein}{0}{70}{70}{-220}{0}
\caption{
A 10s R-Band exposure of V4332 Sgr obtained on 
22 March 1994 at the 1.8-m Perkins telescope with the Imaging Fabry
Perot Spectrograph in direct mode. The field is approximately 
$6\arcmin$ on a side. North is up and West is to the right. }
\end{figure}

\begin{figure}
\plotfiddle{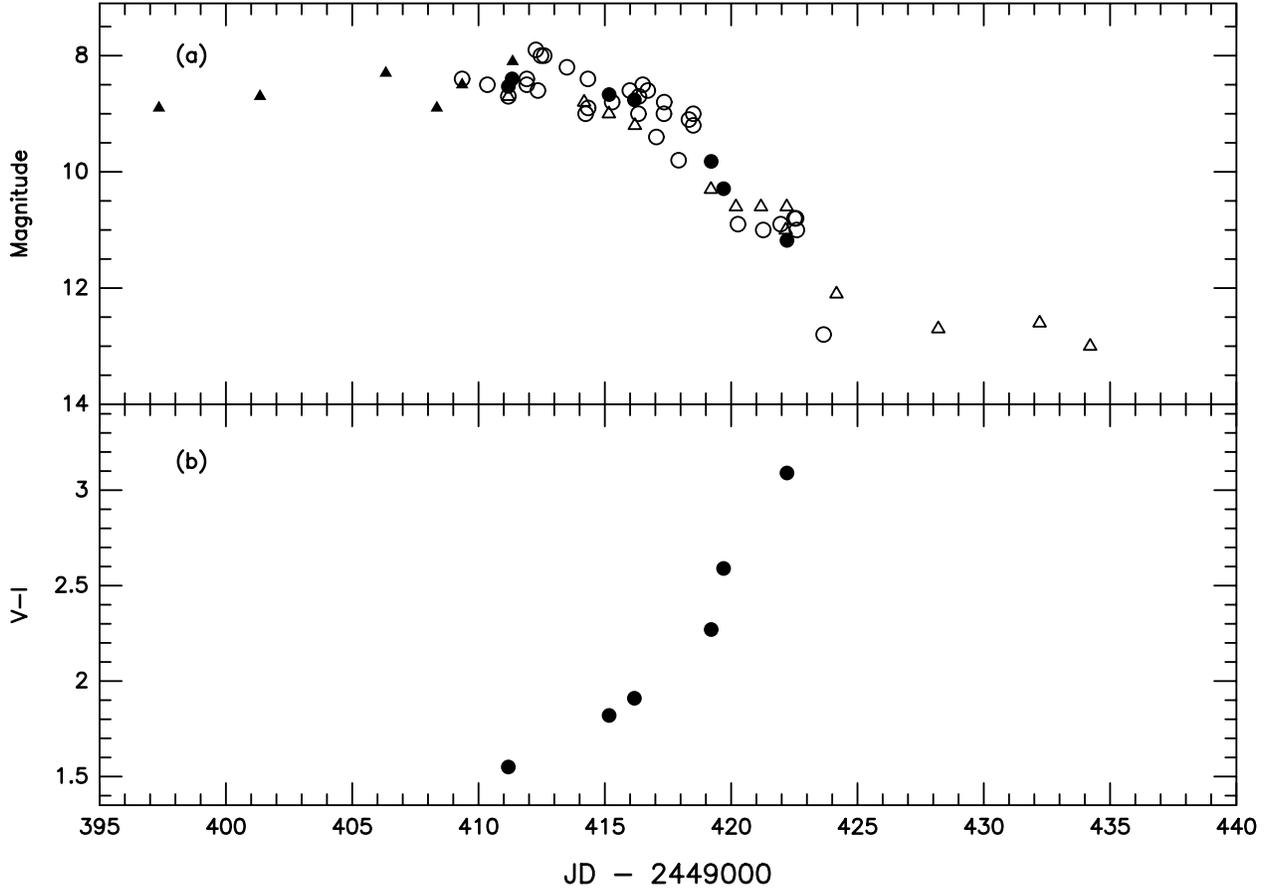}{5.0truein}{0}{70}{70}{-280}{0}
\caption{
Light Curve of V4332 Sgr during outburst. The 
upper figure (a) shows the evolution of V4332 Sgr in the 
visible. In this figure, filled triangles correspond to photographic 
measurements, open triangles correspond to visual estimates, open circles 
represent visual magnitude estimates, and filled circles correspond to 
photoelectric photometry. The lower figure (b) presents the $V-I$ light 
curve over the same period of time where the filled circles represent
photoelectric photometry as in (a). The 1$\sigma$ uncertainties in the 
photoelectric photometry are smaller than the size of the symbols. 
The maximum is roughly at JD 2,449,412, corresponding to 1994 February 28. 
See \S3 for further details. }
\end{figure}

\begin{figure}
\plotfiddle{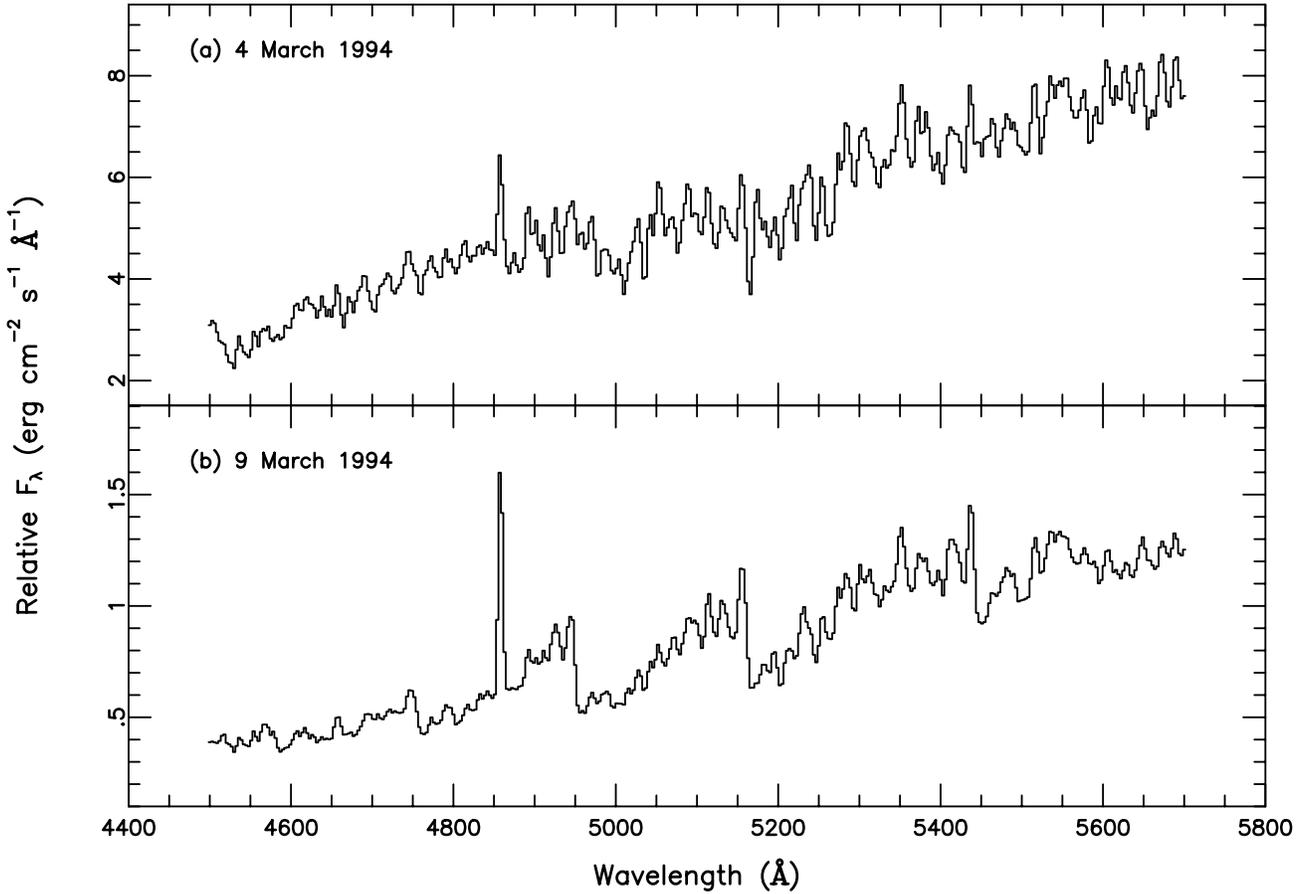}{5.0truein}{0}{70}{70}{-280}{0}
\caption{
March 4 and 9 spectra obtained at the Perkins 
1.8-m telescope at Lowell Observatory. The upper spectrum (a) is from March 4 
and shows relatively little TiO absorption, except for the band at $\lambda$ 
4954 \AA. Also present are numerous metallic absorption features. 
The structure in this spectrum is real and based upon weak TiO absorption and 
the width of the metal lines is classified as K3-4 III-I. The lower 
spectrum (b) shown is from March 9 and shows a remarkable increase in TiO 
absorption bands from (a) above, as well as a significant decrease in flux
(the flux difference between the two spectra is real). 
It is classified as M3 III-I.}
\end{figure}

\begin{figure}
\plotfiddle{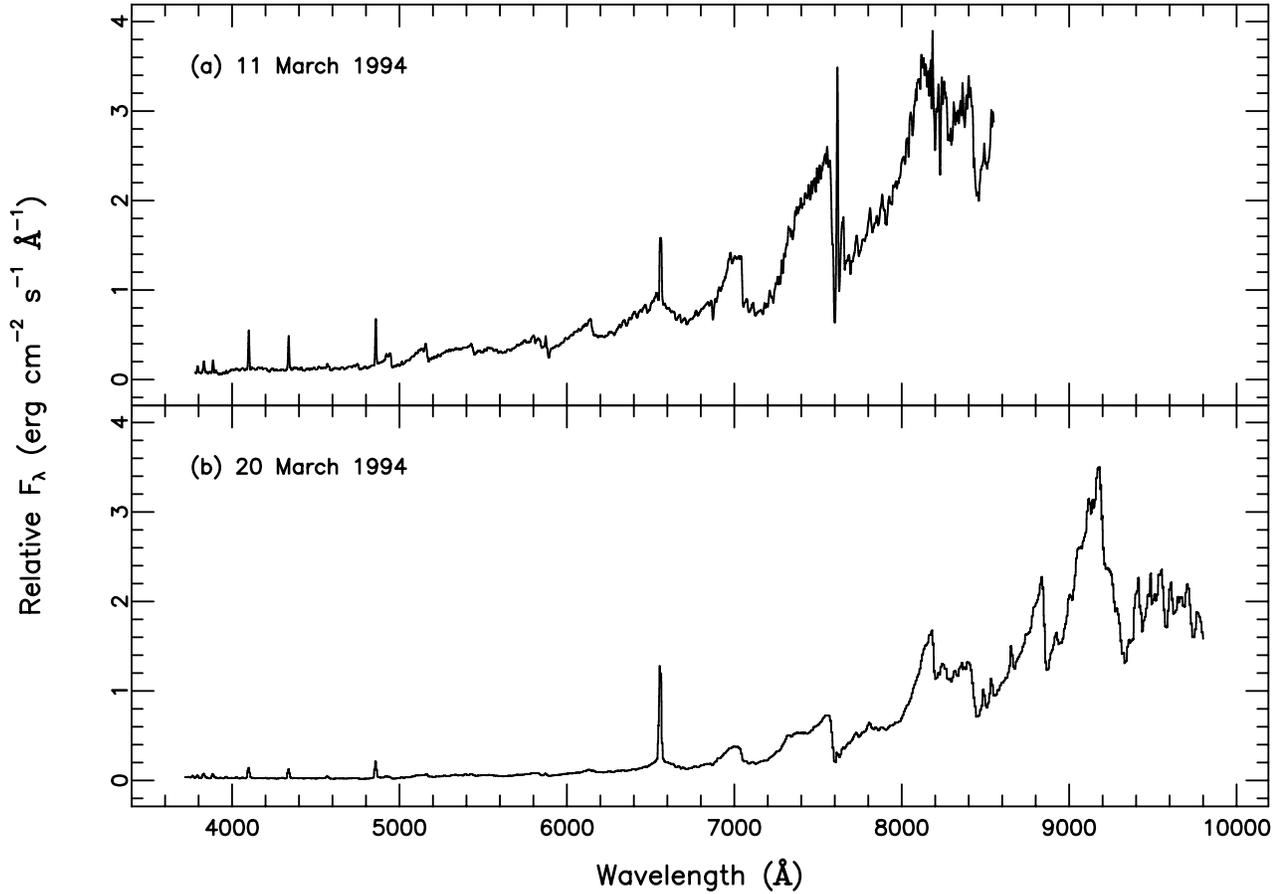}{5.0truein}{0}{70}{70}{-280}{0}
\caption{
March 11 and 20 spectra obtained at the 3.6-m 
ESO telescope at La Silla. The upper spectrum (a) is from March 11 and
shows further evolution towards the red from March 9. The TiO bands 
have increased in strength and the greater blue coverage of this spectrum 
shows the narrow Balmer lines in emission. We classify this spectrum as M5-6
III-II. The lower spectrum (b) is from March 20 and shows weak VO Bands at 
$\lambda\lambda 7400$ \AA\ and 7900 \AA\ characteristic
of class M6.5. Again we find the luminosity class to be of type III-II. In 
addition to the narrow Balmer series present as in (a) above, Mg I 
$\lambda 4571$ is weakly present in emission in both spectra.}
\end{figure}

\begin{figure}
\plotfiddle{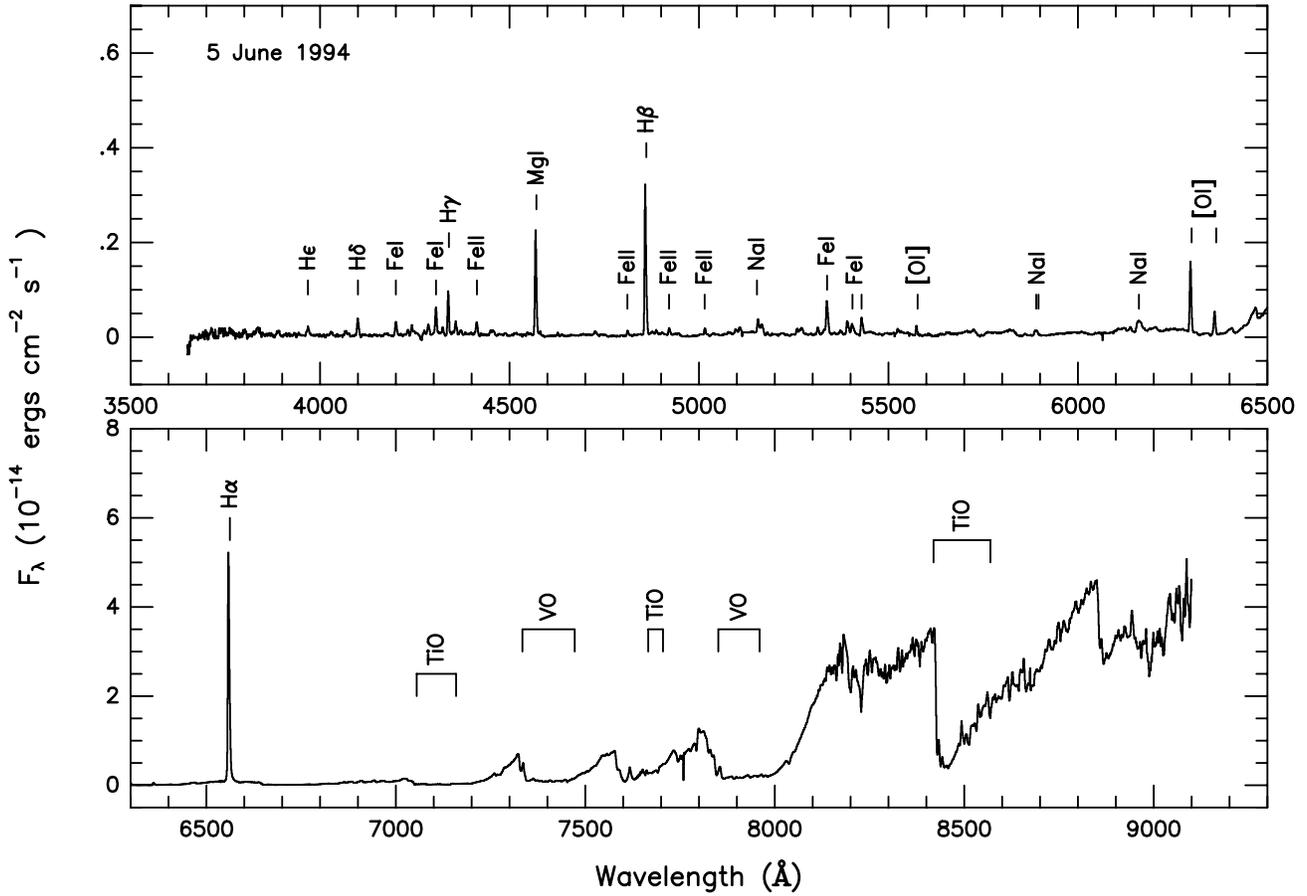}{5.0truein}{0}{70}{70}{-280}{0}
\caption{
Spectrum taken on 1994 June 5/6 at the 4.5-m 
Multiple Mirror Telescope. The bluer component of the spectrum is dominated by 
Balmer emission, Mg I, Fe I, Fe II, Na I, and [O I] (see also Table~3). The 
red component is dominated by strong absorption features, particularly TiO 
and VO. We classify this spectrum as an M8-9 III.}
\end{figure}

\begin{figure}
\plotfiddle{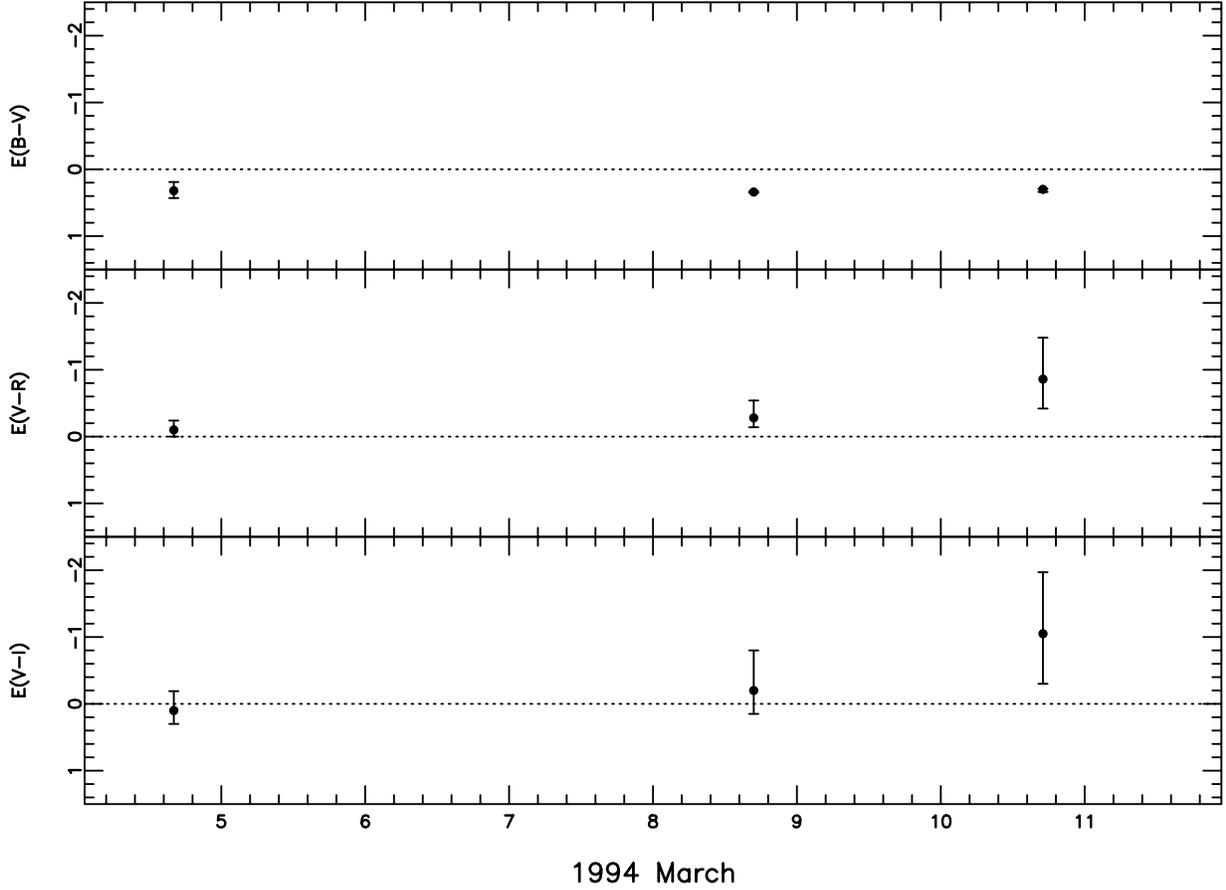}{5.0truein}{0}{70}{70}{-280}{0}
\caption{
The excess color evolution of V4332 Sgr
for the 3 dates on which we have both photometry and spectroscopy. 
We define the color excess as the difference between the observed 
color, based upon published photoelectric photometry, and the expected
color of the object based upon its temperature as derived from our spectra.
The error bars shown correspond to the color difference for an assumed 
uncertainty of one spectral type. Photometric uncertainties are significantly 
less, particularly in the lower two frames. See \S 4 for further details.}
\end{figure}

\begin{figure}
\plotfiddle{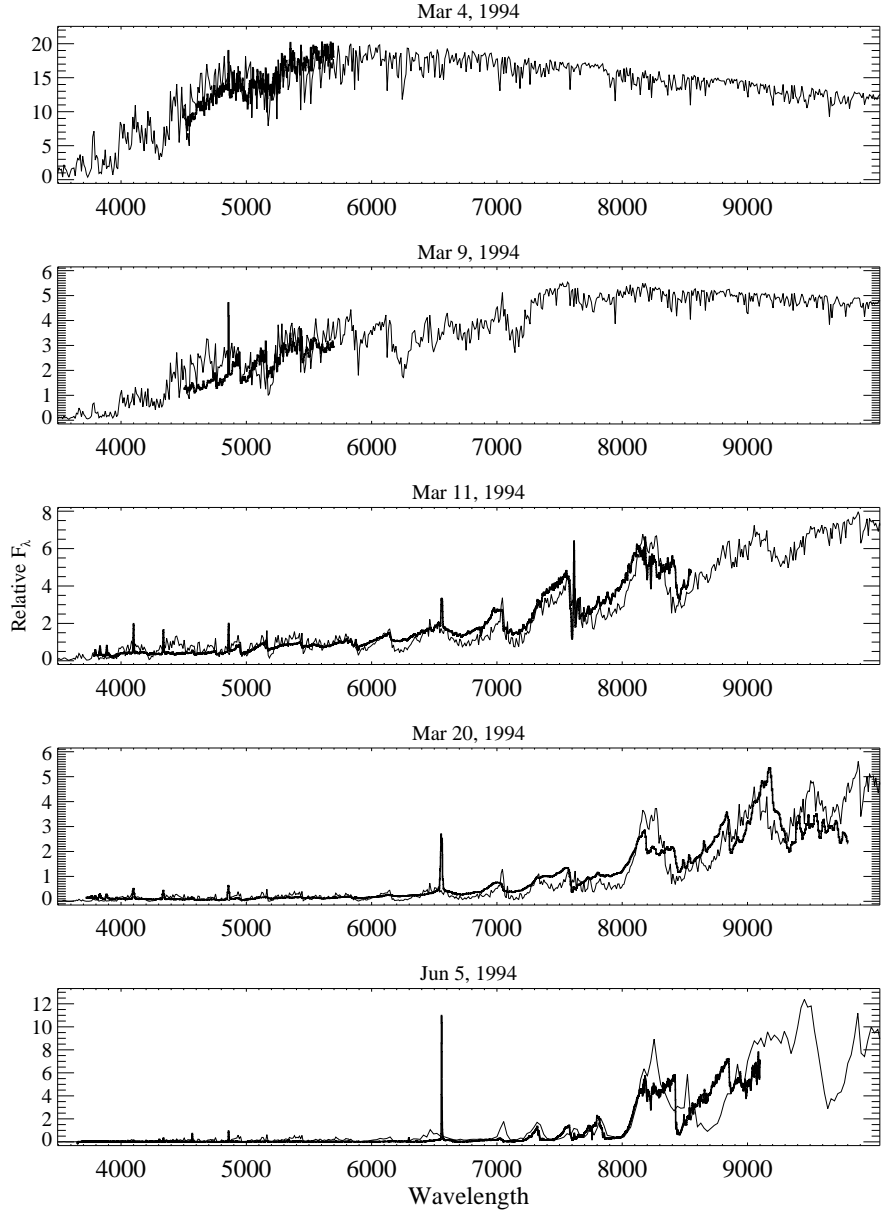}{5.0truein}{0}{70}{70}{-230}{-50}
\caption{
Comparison of the observed reddening corrected
spectra of V4332 Sgr (bold tracing) and the best-fit late-type stellar model
atmosphere as described in the text (see \S 5 for details).  A consistent
spectral range of the model is shown to illustrate the fast spectral evolution
of the V4332 Sgr. In particular, the effective temperatures of the model fits
vary from 4400 K (March 4) to 2300 K (June 5/6). The temperatures corresponding
to these best-fit models are listed in Table~2. }
\end{figure}

\begin{figure}
\plotfiddle{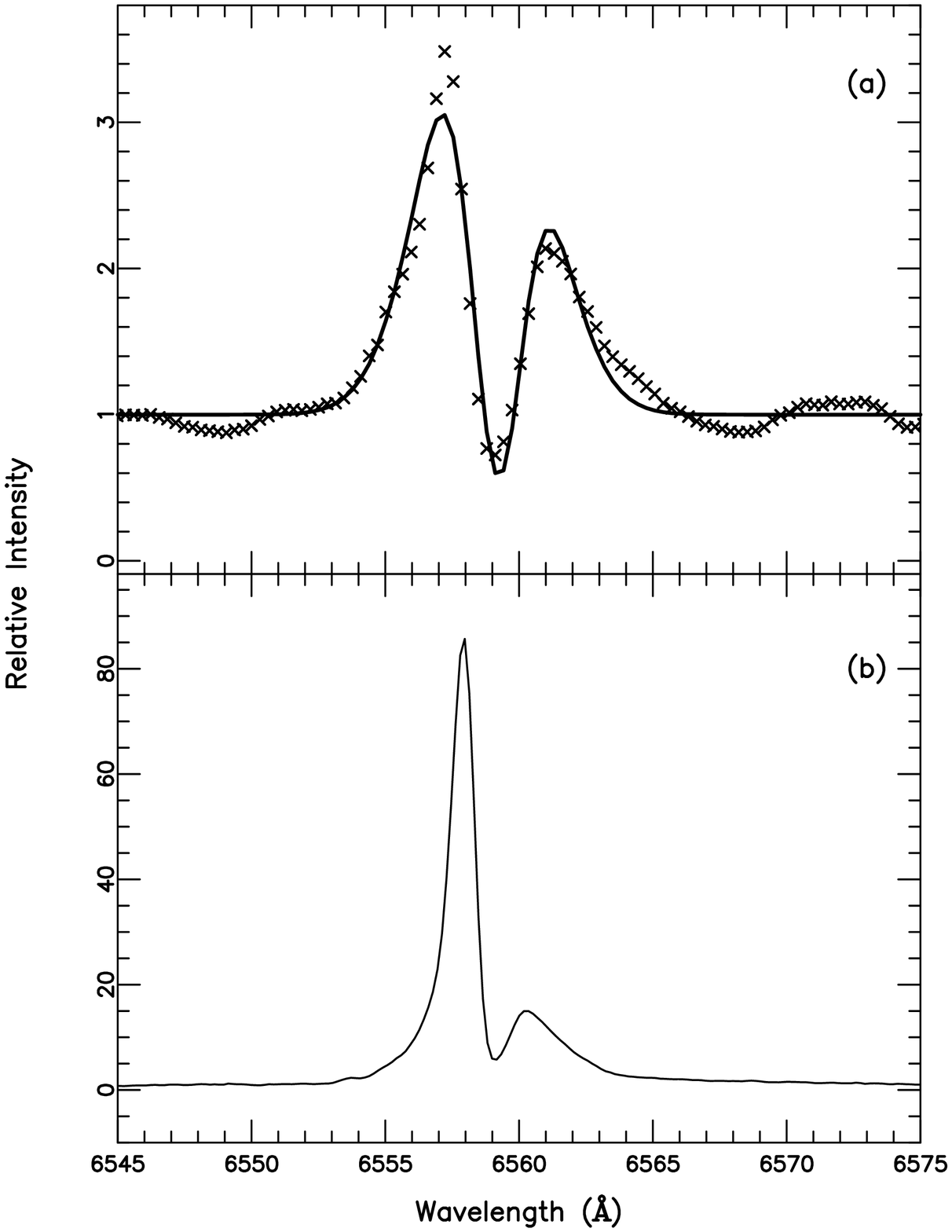}{5.0truein}{0}{70}{70}{-220}{-50}
\caption{
(a) High-resolution spectrum obtained 7 March 1994
at the ESO 3.5-m NTT.  The spectrum, represented by crosses, is centered on the
H$\alpha$ line and has a resolution of 1.4 \AA\ per pixel.  The line fit
represents a simultaneous fit of two gaussian components to the inverse P Cygni
profile; the emission is centered approximately 20 km s$^{-1}$ to the blue of 
the absorption feature.  See \S 6 for further details.  (b) H$\alpha$ line 
profile obtained on 26 April 1994 at the Palomar 1.5-m telescope at a 
resolution of 1 \AA.  Note the dramatic change in the profile width between 
the two dates.}
\end{figure}


\begin{thebibliography}{}

\bibitem[]{} Begelman, M.C. \& Sarazin, C.L. 1986, \apj, 302, L59
\bibitem[]{} Bryan, J. \& Royer, R.E. 1992, \pasp, 104, 179
\bibitem[]{} Duerbeck, H.W. \& Wolf, B. 1977, \aaps, 29, 297
\bibitem[]{} Gallagher, J.S. \& Starrfield, S. 1978, \araa, 16, 171
\bibitem[]{} Gilmore, A.C., Kilmartin, P.M., Sakurai, I. \& Jones, 
  A.F.A. 1994, \iaucirc, 5943
\bibitem[]{} Hauschildt, P.H. \& Baron, E. 1998, Journal of Computational and
Applied Mathematics, {\it in press}
\bibitem[]{} Hauschildt, P.H., Allard, F., \& Baron, E. 1999, \apj, 512, 377
\bibitem[]{} Hayashi, S.S. \& Yamamoto, Y. 1994, \iaucirc, 5942
\bibitem[]{} Iben, I. \& Tutkov, A.V.  1992, \apj, 389, 369
\bibitem[]{} Johnson, H.L. 1966, \araa, 4, 193
\bibitem[]{} Keenan, F.P., Aller, L.H., Hyung, S., \& Brown, P.J.F. 1995, 
  \pasp, 107, 148
\bibitem[]{} Kenyon, S.J. 1986, The Symbiotic Stars, Cambridge University
  Press
\bibitem[]{} Mould, J. et al. 1990, \apj, 353, L35
\bibitem[]{} Partridge, H. \& Schwenke, D. W. 1997, J. Chem. Phys., 106, 4618
\bibitem[]{} Pogge, R.W., Atwood, B., Byard, P.L., O'Brien, T.P., 
  Peterson, B.M., Lame, N.J., \& Baldwin, J.A. 1995, \pasp, 107, 1226
\bibitem[]{} Rich, R.M., Mould, J., Picard, A., Frogel, J.A., 
  \& Davies, R. 1989, \apj, 341, L51
\bibitem[]{} Sato, H., Sakurai, Y., Beers, T., Brandner, W., Lehmann, T., 
  Duerbeck, H. W., Della Valle, M., \& Smette, A., \iaucirc, 6051
\bibitem[]{} Sharov, A.S. 1990, {\it Sov. Astron. Lett.} 16, 85
\bibitem[]{} Sharov, A.S. 1993, {\it Sov. Astron. Lett.} 19, 33
\bibitem[]{} Tomaney, A.B. \& Shafter, A.W. 1992, \apjs, 81, 683
\bibitem[]{} Tomaney, A.B., Rich, R.M., Wagner, R.M., \& Della Valle, M.
  1994, \iaucirc, 5949
\bibitem[]{} Turnshek, D.E., Turnshek, D.A., Craine, E.R., \& 
  Boeshaar, P.C. 1985, An Atlas of Digital Spectra of Cool Stars, Western 
  Research Company
\bibitem[]{} Wagner, R.M. 1994, \iaucirc, 5944
\bibitem[]{} Xu, Y., McCray, R., Oliva, E., \& Randich, S. 1992, \apj, 386, 181

\end{thebibliography}
\end{document}